\RequirePackage[2020-02-02]{latexrelease}
\documentclass[twocolumn,tighten]{aastex631mod}
\usepackage{xcolor}

\newcommand{\kms}{km~s$^{-1}$}
\newcommand{\kmsMpc}{km~s$^{-1}$\,Mpc$^{-1}$}
\newcommand{\ho}{\ensuremath{H_{\rm 0}}}
\newcommand{\dmax}{\ensuremath{d_\mathrm{Max}}}
\newcommand\sna{SN\,Ia}

\shorttitle{Roman Core Community Survey White Paper}
\shortauthors{SBF Distances with Roman}

\begin{document}
\title{Gathering Galaxy Distances in Abundance with Roman Wide-Area Data}


\author[0000-0002-5213-3548]{John P. Blakeslee}
\affiliation{NSF's NOIRLab, Tucson, AZ, USA; 
\,\href{mailto:john.blakeslee@noirlab.edu}{john.blakeslee@noirlab.edu}}

\author[0000-0003-2072-384X]{Michele Cantiello}
\affiliation{INAF-Abruzzo, Teramo, Abruzzo, Italy; 
\href{michele.cantiello@inaf.it}{michele.cantiello@inaf.it}}

\author[0000-0002-1437-3786]{Michael J. Hudson}
\affiliation{University of Waterloo, Waterloo, ON, Canada}

\author[0000-0002-8224-1128]{Laura Ferrarese}
\affiliation{NRC Herzberg Astronomy and Astrophysics Research Centre, Victoria, BC, Canada}

\author[0000-0002-3870-1537]{Nandini Hazra}
\affiliation{Gran Sasso Science Institute, L’Aquila, AQ, Italy}

\author[0000-0001-8762-8906]{Joseph B. Jensen}
\affiliation{Utah Valley University, Orem, UT, USA}

\author[0000-0002-2073-2781]{Eric W.~Peng}
\affiliation{NSF's NOIRLab, Tucson, AZ, USA}

\author[0000-0002-5577-7023]{Gabriella Raimondo}
\affiliation{INAF-Abruzzo, Teramo, Abruzzo, Italy}


\begin{abstract}%
\noindent
The extragalactic distance scale is fundamental to our understanding of astrophysics and cosmology. In recent years, the surface brightness fluctuation (SBF) method, applied in the near-IR, has proven especially powerful for measuring galaxy distances, first with HST and now with a new JWST program to calibrate the method directly from the tip of the red giant branch (TRGB). So far, however, the distances from space have been gathered slowly, one or two at a time. With the Roman Space Telescope, we have the opportunity to measure uniformly high-quality SBF distances to thousands of galaxies out to hundreds of Mpc. 
The impact of these data on cosmology and galaxy studies depends on the specifics of the survey, including the filter selection, exposure depth, and (especially) the sky coverage. 
While the baseline HLWAS survey in four filters plus the grism would yield useful data, the impact would be limited by the relatively small area. A more optimal approach would concentrate on the most efficient passband (F146), adopt an exposure time sufficient to measure good quality distances well out into the Hubble flow ($z\gtrsim0.03$), and then maximize the sky coverage within the total time constraints. Grism observations over the same area can provide the needed information on redshifts and spectral energy distributions for compact sources, while colors for larger objects can be obtained from lower resolution surveys.
The proposed plan will enable accurate determination of the physical properties of thousands of nearby galaxies, an independent measure of the Hubble constant $H_0$ with negligible statistical error, and competitive constraints on  $S_8{\,=\,}\sigma_8(\Omega_m/0.3)^{0.5}$. The resulting data set will be a phenomenal resource for a wide range of studies in astrophysics and cosmology.
\\[4pt]
\noindent\textbf{Roman Core Community Survey:} High Latitude Wide Area Survey\\[3pt]
\textbf{Scientific Categories:\,} galaxies --
large-scale structure of the universe:~cosmological parameters 
\end{abstract}

\section{Measuring the Universe}
\label{section:intro}

The fields of extragalactic astronomy and observational cosmology began in earnest a century ago with the identification of Cepheid variables in spiral nebulae \citep{Hubble1925}, leading to the discovery of the expanding universe \citep{Lemaitre1927,Hubble1929}. The distances at the time were crude: Hubble's data set had four galaxies in the Virgo cluster, which he took to be at 2~Mpc. But as the distances improved, so did our understanding of the universe. By the turn of the millennium, 
relative distances from Type Ia supernovae (SN\,Ia), corrected for decline rate, led to the discovery that the universe was not only expanding but accelerating \citep{Riess1998,Perlmutter1999}, while the  Cepheid-based calibration of a variety of distance indicators constrained the Hubble constant $H_0$ to within 10\% \citep{Ferrarese2000,Freedman2001}.

Soon afterwards, results from WMAP on the cosmic microwave background (CMB) seemed to confirm the $\Lambda$CDM ``concordance cosmology" with $H_0 \approx 70$ \kmsMpc\ 
\citep{Bennett2003}.
But cracks in this edifice began to show about a decade later when the early universe CMB measurements and the late-universe SN\,Ia distances had improved enough that the margins of error on $H_0$ no longer overlapped.
This ``Hubble tension" has grown progressively worse and now exceeds $5\sigma$ in significance, with the latest SN\,Ia-Cepheid analysis giving $H_0 =73.0\pm1.0$ \citep{Riess2022},
as compared to $H_0 =67.4\pm0.5$ \citep{Planck2020} predicted from analysis of the CMB
\citep[see the review by][]{intertwined2022}. This may point towards physics beyond the standard cosmological model \citep[e.g.,][]{DiValentino2021}, but another analysis of the SN\,Ia distances, using tip of the red giant branch (TRGB) distances for calibration, finds $\ho=69.8\pm1.8$, consistent with the CMB prediction \citep{Freedman2021}. Clearly, we need other independent routes, not involving Cepheids or SN\,Ia, to $H_0$ in the local universe.~~~~ 

One promising path involves the surface brightness fluctuations (SBF) method calibrated from the TRGB. A recent work reports $H_0 =73.3\pm2.5$ from 63 SBF distances out to 100 Mpc observed with WFC3/IR on Hubble \citep[][]{blake21,jensen21}. The calibration was mainly based on Cepheids, a precarious scaffolding for SBF, which works best for early-type galaxies. However, a new JWST Cycle 2 program will establish a firmer footing for the method using NIRCam to measure TRGB and SBF distances for an optimally selected set of 14 nearby ellipticals. This will enable a fully independent value of \ho\ with a precision rivaling that of \sna\ calibrated via Cepheids. To bring this approach to full fruition will require hundreds of SBF distances spread across the sky and reaching to at least $z{\,\sim\,}0.03$, where bulk flows are thought to be negligible.

\section{Galaxy Properties and Dark Matter}

Of course, reliable distances tell us about more than ``just'' cosmology. They are essential for converting observed properties  
into physical quantities such as size, mass, luminosity and energy. Yet, except for very nearby, resolved systems, they are notoriously difficult to estimate, 
with occasional “factor-of-two” controversies \citep[e.g.,][]{schweizer08,trujillo19}. In their review of black hole scaling relations, \citet{kormendyho13} point out that distance errors dominate the uncertainty for many black hole mass estimates, even though authors neglect it in their final quoted errors. And if distance is a major source of error in black hole mass, which scales linearly with $d$, it is almost always the dominant error for galaxy luminosity, which scales as $d^2.$ This can have important implications for understanding the nature of some systems.

To take one example, the diffuse galaxy NGC\,1052--DF2 was claimed by \citet{vdk18} to be devoid of dark matter, based on an SBF distance of $\sim\,$20~Mpc. A subsequent study argued that the galaxy had a relatively normal dark matter content based on a distance of 13 Mpc, estimated mainly from the globular clusters \citep{trujillo19}. Thus, the interpretation was wildly different, depending on the distance.  A subsequent measurement of the tip of the red giant branch (TRGB) yielded $d$ = 22.1 $\pm$ 1.2 Mpc \citep{shen21}, consistent with the SBF distance  $d$ = 20.4 $\pm$ 2.0 Mpc \citep{blake18}. 

The remarkable thing is that the TRGB distance used 40 HST orbits, while the SBF distance was based on a single orbit. Unlike other precision methods (Cepheids, TRGB, \sna, masers), SBF requires only modest depth and no monitoring. Besides the WFC3/IR \ho\ study mentioned above, SBF has been used with HST to study the structure of nearby galaxy clusters \citep{Mei2007,blake09}, convert the observed “shadow” of the M\,87 supermassive black hole into a physical size and mass \citep{ehtVI}, measure the most precise distance to the host galaxy of the gravitational wave source GW170817 \citep{cantiello18}, and explore possible nonlinearity in the \sna\ peak luminosity versus decline rate \citep{Garnavich2022}.

However, with HST, optical and near-IR SBF measurements have accumulated one pointing at a time in the course of a dozen GO programs over two decades. With JWST, the exceptional imaging capabilities make it possible to establish a rock-solid calibration for the method and extend the range to twice that reached with HST. But in most cases, the JWST/NIRCam field of view also only accommodates one target at a time. Consequently, it is best suited for determining precise distances for specific individual targets, rather than ``harvesting" SBF distances in the thousands. For this purpose, we require the Roman Observatory, guided by a well-defined wide-area survey observing strategy.



\section{``Notional" Roman HLWAS Numbers}

Roman Observatory, with its Wide Field Instrument, presents unprecedented opportunities for  distance studies using SBF to constrain cosmology and galaxy properties. Although bands at the red end of the optical spectrum minimize the intrinsic scatter in the SBF method, near-IR bands like $J,H,K$ offer several advantages. The fluctuations themselves are inherently brighter in the near-IR, with at least ten times higher amplitude in $K$ than $I$ \citep[e.g.,][]{jensen98}. The near-IR also gives a much more favorable contrast compared to globular clusters, the main contaminant in measuring SBF distances for giant ellipticals (as discussed below). Finally, the effects of residual dust contamination is greatly reduced. For all these reasons, recent space-based SBF studies have focused on the near-IR.

To illustrate, the ACS Virgo and Fornax Cluster Surveys yielded SBF distances for over 130 galaxies in these two clusters, with one HST orbit dedicated to each galaxy \citep{blake09}. This enabled an exquisite calibration of the stellar population dependence of the method, along with a precise value of the relative distance of the clusters.  With Roman/WFI we can measure a similar number of galaxies in the more distant Coma cluster with only $\sim\,$7 pointings, and with a similar exposure time per pointing (i.e., $\sim\,$5\% of the total time) because of the brightness of the SBF signal in the near-IR and the wide area of WFI instrument.

Roman's High Latitude Wide Area Survey (HLWAS) promises to revolutionize this field.
For a given survey specification, we wish to quantify both the maximum distance \dmax\ to which SBF measurements can be made and the number of reliable galaxy distances.
For giant ellipticals, contamination of the power spectrum by globular clusters (GCs) is the main limiting factor \citep{moresco22}. Thus, \dmax\ is the distance to which the GCs can be detected (at $5\sigma$) and removed to a faint enough limit, and the residual contamination reliably estimated, so that the uncertainty in the correction drops below the intrinsic scatter in the method.~~~ 
%

For instance, in the $I$ band, where the peak of the GC luminosity function (GCLF) is at $M_I\approx-8.0$ AB mag, accurate SBF measurements require detecting and removing sources to +0.5~mag fainter than the GCLF peak, or $M_I\approx-7.5$ AB. In the $H$ band, the same relative amount of GCs contamination can be reached for a detection limit relative to the GCLF peak of about $-$1.0 mag 
\citep[i.e., 1 mag \textit{brighter} than the peak;][]{jensen21}. 
We estimate the near-IR GCLF peak absolute magnitudes based on \citet{nantais06}, converted to AB.
In addition, because they are projected against the bright galaxy background, we have found that the $5\sigma$ detection limit for the GCs is on average $\sim0.5$ mag brighter than for isolated point sources. With these assumptions, and an adopted space density of early-type galaxies, we can estimate \dmax\ and the expected yield of SBF distances for a given survey design.

To estimate the number of early-type galaxies suitable for SBF measurement, we first use the 2MASS Redshift Survey (2MRS), which contains 43,000 galaxies with $K_s<11.75$ mag covering  91\% of the sky \citep{huchra12}. For reference, an $L^*$ galaxy will be included in the 2MRS for $d\lesssim135$ Mpc. The 2MRS includes the morphological $T$-type, and we select only galaxies with $T\leq-1$, indicating early-type, and with an absolute magnitude $K_s<-20$ mag, estimated from the redshift. To improve completeness for $L^*$ galaxies at distances of interest, we repeat the calculations using the 2M++ catalog \citep{2mpp}, which augments the 2MRS with deeper data over much of the sky. The 2M++ does not provide $T$-type, so we assume the same early-type percentage (38\%) as in the 2MRS for our adopted absolute magnitude limit. Finally, we assume a flat distribution of galaxies on the sky, as the eventual location of the HLWAS is unknown.

For the observational details, we first adopt the ``notional" HLWAS as specified online 
for the F106, F129, and F158 bands. We omit the lower efficiency F184 band and instead 
report expectations for the broad, high-throughput F146 band with a similar exposure time. 
The first four rows of Table~1 give the estimated maximum distances and expected numbers of suitable SBF targets, along with all the input assumptions, for each of these bands with the notional 2000~deg$^{2}$ area coverage.

The numbers are impressive. Even in the $J$ band, $\sim\,$300 galaxy distances reaching out to $\sim\,$110 Mpc would significantly reduce the statistical error on the present value of \ho\ from SBF. However, limiting the coverage to $\sim\,$5\% of the sky in a single direction opens the possibility for systematic errors due to flow motions, clustering, and potentially other forms of cosmic variance. In the final section, we propose a more optimal survey design for constraining cosmological parameters.

\section{Optimizing the Survey Strategy}

There are two prominent “tensions” in cosmology: the Hubble tension, discussed above, and
the $S_8$ tension, where $S_8\equiv\sigma_8(\Omega_m/0.3)^{0.5}$ quantifies the level of
matter inhomogeneity in the universe.  In both cases, the tension is between the
predicted value extrapolated from the CMB (assuming $\Lambda$CDM) and the value measured
in the local universe. These are the biggest problems in cosmology today; they may well
have a common underlying explanation \citep[e.g.,][]{intertwined2022}.

The strongest evidence for the $S_8$ tension comes from weak lensing
\citep[e.g.,][]{Amon2023}, but results from peculiar velocities
\citep{Boruah2020,Said2020} point in the same direction, with larger
uncertainties. Different weak lensing surveys may share common systematics, such as
intrinsic alignments; thus, it is critical to confirm the $S_8$ tension using diverse
methods. For peculiar velocity studies, it is most important to maximize the sample size
and volume, while keeping the distance errors at a level comparable to the peculiar
velocities themselves. SBF distance errors are typically 5-6\%, or $\sim\,$400 \kms\ at
100 Mpc. This is vastly better than typical errors of 20-25\% from other galaxy-based
methods such as the Fundamental Plane or Tully-Fisher.

The final row of Table~1 shows the expected results for an illustrative
wide survey covering a quarter of the sky in F146, the most efficient of the WFI bands,
to a depth similar to those envisioned for the four filters in the notional HLWAS shown
in the top part of the table. This strategy would deliver a peculiar velocity sample with
2500 to 5000 galaxies, reaching out to 150 Mpc with 6\% error. Such a data set would be
unprecedented in its combination of precision and sample size. \sna\ distances have
similar precision but are much rarer, with only about 500 available within the same
volume. Following the methodology of \citet{Boruah2020}, we predict uncertainties on
$S_8$ from this hypothetical SBF distance sample to be below 2\%, limited by cosmic
variance uncertainties in the density field. This is competitive with the best current
weak lensing results.

Our proposal then is to cover as wide an area as possible using Roman's most efficient filter to a depth where detector noise becomes negligible. Analysis of~the S/N curve for F146 suggests
$\sim\,$100\,s per exposure; a 3-point dither pattern then gives 5~min per pointing. 
Comprehensive grism data would provide redshifts and SEDs for compact sources; 
colors for larger objects can be obtained from ground-based optical and near-IR surveys.
While a $\pi$-sterradian HLWAS may be overly ambitious, even coverage of 10\% of the sky 
would greatly reduce systematics from cosmic variance.
This approach would yield high-quality distances for thousands of galaxies, an independent measure of $H_0$ with negligible statistical error, and competitive constraints on $S_8$. The data set would be an enormously rich resource for a wide range of studies in astrophysics and cosmology.


\begin{table*}
~\vspace{0.5cm}
\caption{Number of Candidates for SBF analysis}
\centering
\begin{tabular}{lccccccc} \hline \hline
 (1) & (2)  & (3) &  (4)  &  (5) &  (6) &  (7) &  (8)   \\
 Band & ~~Depth$^{(a)}$  & GCLF &  GCLF       & $(m{-}M)$ & $d_\mathrm{Max}$ & targets & targets   \\
      &  (mag) & peak$^{(b)}$ &  offset$^{(c)}$ & (mag) &     (Mpc)  & 2MRS$^{(d)}$ & 2M++$^{(e)}$  \\
\hline 
%
\multicolumn{8}{c}{~~~Baseline ~2000 deg$^2$~ HLWAS $+$ F146 } \\ 
F106 (${\sim\,}Y$) & 26.4 & $-$8.1 & $+$0.0 &  34.5 & 80  &  180 &  290  \\  
F129 (${\sim\,}J$) & 26.5 & $-$8.2 & $-$0.5 &  35.2 & 110 &  280 &  480  \\  
F158 (${\sim\,}H$) & 26.4 & $-$8.5 & $-$1.1 &  35.9 & 150 &  400 &  750  \\ 
F146 (${Y}{J}{H+}$) & 26.7 & $-$8.4 & $-$1.0 &  36.1 & 165 &  500 & 950  \\[2pt]
\hline 
\multicolumn{8}{c}{F146 $\pi$-steradian survey$^{(f)}$ } \\ 
F146 (${Y}{J}{H+}$) & 26.7   & $-$8.4 & $-$1.0 &  36.1 & 165 &  2500 & 4800 \\[1pt]  
\hline  \hline
\multicolumn{8}{l}{
  \begin{minipage}{11.7cm}%
  \vspace{6pt}
\scriptsize $(a)$ Nominal $5\sigma$ point-source depth accounting for host galaxy background.\\[3pt]
\scriptsize $(b)$ Peak, or turnover, magnitude (AB) of the GCLF in each band \\[3pt]
\scriptsize $(c)$ Offset of the required detection limit with respect to the GCLF peak (see text).\\[3pt]
\scriptsize $(d)$ Estimated number of suitable SBF targets based on 2MRS \citep{huchra12}.\\[3pt]
\scriptsize $(e)$ Estimated number of suitable SBF targets based on 2M++ \citep{2mpp}.\\[3pt]
\scriptsize $(f)$ Illustrative survey covering a quarter of the sky, focusing on the high-throughput F146 filter.\\[5pt]
    \end{minipage}
    }\\
\end{tabular}
  \newline       
\end{table*}

\bibliography{sbfroman}{}
\bibliographystyle{aasjournal}

\end{document}